%% file: main.tex
\documentclass{article}
\usepackage{authblk} 
\usepackage{arxiv}
\usepackage[utf8]{inputenc} 
\usepackage[T1]{fontenc}    
\usepackage{graphicx}
\usepackage{amsmath}
\usepackage{bm}
\usepackage{hyperref}
\usepackage{xcolor}
\hypersetup{
    colorlinks=true,
    linkcolor=blue,        
    filecolor=magenta,     
    urlcolor=cyan,         
    citecolor=purple,         
    pdftitle={},
    pdfpagemode=FullScreen,
}
\usepackage{url}            
\usepackage{booktabs}       
\usepackage{amsfonts}       
\usepackage{nicefrac}       
\usepackage{microtype}      
\usepackage{float}
\floatstyle{plaintop}
\restylefloat{table}
\usepackage{algorithm}
\usepackage{algpseudocode}
\usepackage{setspace}
\usepackage{lipsum}
\usepackage{subcaption}
\usepackage[numbers,sort&compress]{natbib}

\newcommand{\red}[1]{\textcolor{black}{#1}}   

\def\figSizeSingle{0.55}

\title{A posteriori closure of turbulence models: are symmetries preserved?}

\author[1,2]{\textbf{André Freitas}$^*$}
\author[2]{\textbf{Kiwon Um}}
\author[3]{\textbf{Mathieu Desbrun}}
\author[1]{\textbf{Michele Buzzicotti}}
\author[1]{\textbf{Luca Biferale}}

\affil[1]{Dept. Physics and INFN, University of Rome "Tor Vergata" and INFN, Italy}
\affil[2]{LTCI, Télécom Paris, IP Paris, France}
\affil[3]{LIX, Inria and École Polytechnique, IP Paris, France}

\affil[*]{Corresponding author: \texttt{andre.freitas@roma2.infn.it}}

\begin{document}
\maketitle
\begin{abstract}
Turbulence modeling remains a longstanding challenge in fluid dynamics. Recent advances in data-driven methods have led to a surge of novel approaches aimed at addressing this problem. This work builds upon our recent work [Phys. Rev. Fluids 10, 044602 (2025)], where we introduced a new closure for a shell model of turbulence using an a posteriori (or solver-in-the-loop) approach. Unlike most deep learning-based models, our method explicitly incorporates physical equations into the neural network framework, ensuring that the closure remains constrained by the underlying physics benefiting from enhanced stability and generalizability.
In this paper, we further analyze the learned closure, probing its capabilities and limitations. \red{In particular, we look at joint probability density functions between resolved and unresolved variables, as well as the scale invariance of multipliers (ratios between adjacent shells) within the inertial range. Although our model excels in reproducing high-order statistical moments, it breaks this known symmetry near the cutoff, indicating a fundamental limitation.} We discuss the implications of these findings for subgrid-scale modeling in 3D turbulence and outline directions for future research.
\end{abstract}

\input{sections_V2/intro}

\input{sections_V2/sec2}

\input{sections_V2/method}
\input{sections_V2/sec3}
\input{sections_V2/sec4}
\input{sections_V2/acknowledgements}
\pagebreak
\clearpage
\bibliographystyle{unsrt}
\bibliography{references}  

\clearpage
\appendix
\input{sections_V2/appendix}

\end{document}

%% file: sections_V2/intro.tex
\section{Introduction}
{\color{black}
Turbulence is ubiquitous in nature, from the coffee we stir in the morning to the primordial states of our universe~\cite{frisch}. Despite its ubiquity, understanding and modeling turbulence remains one of the most challenging tasks in classical physics. While we have known its governing equations for over a century~\cite{voyage2011}, the full phenomenology of turbulence remains elusive~\cite{sreeni25, falkovich2006lessons}. Simulating turbulence numerically is also extremely computationally demanding~\cite{dns_moin, kim_leonard_review}. The degrees of freedom scale as a power law of the Reynolds number, $Re = UL/\nu$ (where \(U\) is a characteristic velocity, \(L\) the integral length scale, and \(\nu\) the kinematic viscosity), which means that studying very high Reynolds number turbulence remains impossible for the most part. One approach to reduce the degrees of freedom is Large Eddy Simulation (LES)~\cite{les_review}, where a filter is applied, typically within the inertial range, to resolve only the largest, most energy-containing eddies. However, to close the LES system, the effects of unresolved subgrid scales must be modeled. This constitutes the LES modeling problem. This problem is often addressed phenomenologically by introducing an eddy diffusivity term to stabilize the system~\cite{smag63}. This solution, albeit useful for practical application, fails to capture the correct subgrid scale statistics. Moreover, although decades of research have been spent in LES modeling~\cite{smag63,lilly1967proposed,kraichnan76,new_directions_LES_1996,Biferale_AAM_Parisi,duraisamy2019,Buzzicotti_2023}, there are still fundamental open questions in the field. Can a subgrid-scale closure be formulated to accurately capture the statistics of the resolved scales, especially in high Reynolds number flows where intermittency~\cite{int1993prl} plays a crucial role? In turbulence, velocity fields are not statistically self-similar due to intermittency, making it essential for closures to account for the multifractal nature of the subgrid scales~\cite{parisi1985predictability}.

Recently, deep learning (DL)~\cite{Goodfellow-et-al-2016} approaches have gained attention for addressing the LES modeling problem~\cite{duraisamy2019}. DL excels at approximating functions with unknown functional forms~\cite{cybenko1989approximation, hornik1991approximation}. By learning from data, DL methods can, albeit in a black-box manner~\cite{mythos_interpretability}, accurately approximate high-dimensional functions~\cite{bengio2007scaling}.} To investigate the applicability of DL in this context, we use shell models of turbulence~\cite{biferale2003shell} as our playground. Shell models offer a simplified representation of 3D homogeneous isotropic turbulence (HIT), providing a tractable alternative to the full Navier-Stokes equations. Despite their simplicity, shell models preserve key turbulence phenomena, such as energy cascade, intermittency, anomalous scaling, and multifractal energy dissipation.

Our primary focus is on the ability of subgrid-scale (SGS) models to accurately capture extreme events and rare fluctuations, which demand high precision, extensive statistical sampling, and long inertial ranges. Learning SGS models from data requires long temporal evolutions, presenting a fundamental challenge for LES of three-dimensional turbulence. This challenge is currently best addressed through shell models. Although LES of three-dimensional HIT remains the ideal test case, existing machine learning-based LES tools are hindered by their low resolution and the proximity of the cutoff to the forcing scales, preventing the emergence of anomalous behavior. Shell models, in contrast, provide a more suitable framework for rigorously testing the effectiveness of LES models under conditions of strong intermittency and for assessing whether DL approaches can improve the tracking of extreme fluctuations over time.
Furthermore, higher-order statistics--particularly of fourth order and beyond--naturally involve non-local interactions in Fourier space, implying that SGS fluctuations can influence the resolved scales in non-trivial ways. This observation is critical when aiming for accuracy beyond second-order moments, which are typically the focus of LES validation. Shell models thus offer a controlled environment to probe whether a closure model can correctly capture such high-order interactions.

Several previous studies have explored closures within the shell model framework. Biferale et al.~\cite{Biferale_AAM_Parisi} initially approached the closure problem phenomenologically, using Kolmogorov multipliers. More recently, Ortali et al.~\cite{Ortali22} employed deep learning to learn the closure, while Domingues Lemos~\cite{JDL24, JDL25} integrated probabilistic machine learning tools within the hidden symmetric framework of shell models~\cite{HS_21} to address the closure. Building on these developments, we have also contributed to the field~\cite{freitas25} through a solver-in-the-loop approach~\cite{um2020sol} based on the differentiable physics paradigm. This method unrolls the system in time during training, allowing the neural network to interact with a numerical solver for the governing equations over an arbitrary number of time steps before the loss function is computed. Rather than attempting to match the instantaneous, \emph{a priori} evolution of the ground truth, we adopt an \emph{a posteriori} approach, where the network learns how its errors propagate through time while receiving more realistic inputs. The induced closure was analyzed by examining high-order moments such as flatness up to the tenth order, demonstrating good agreement with the ground truth and unconditionally stable behavior.

\red{This paper further extends our previous investigation by studying the learned closure in greater detail. While our earlier work~\cite{freitas25} showed that the model reproduces the statistics of the resolved variables with high accuracy, here we test a stricter criterion: whether it can also recover nontrivial correlations between resolved and unresolved variables. To this end, we analyze joint probability density functions and other multi-scale statistics. We also investigate Kolmogorov multipliers~\cite{Kolmogorov_1962}, which are known to exhibit universal, scale-invariant statistics within the inertial range. These multipliers, especially those near the cutoff, intrinsically couple resolved and unresolved variables and therefore provide a sensitive diagnostic of whether the learned closure preserves or breaks this symmetry.}

The remainder of the paper is structured as follows. In Section 2, we introduce the shell model framework and the mathematical formulation of our problem. Section 3 details the solver-in-the-loop approach and the methodology for learning an a posteriori closure. In Section 4, we present our results, focusing on statistical properties and preservation of known symmetries. Finally, we discuss our findings in Section 5 and outline potential directions for future research.

%% file: sections_V2/sec2.tex
\section{Shell models}

Shell models are a class of turbulence models. They model homogeneous isotropic turbulence in Fourier space, considering only sparse coupling between the different shells. We discretize the 1D Fourier space with a geometric progression of shells that are equispaced on a logarithmic lattice, $k_n = k_0\lambda^n$, where one typically uses $\lambda \!=\! 2$ and $k_0 \!=\! 1$. We consider the Sabra shell model~\cite{LvovUnknownTitle1998}, where the set of ordinary differential equations governing the dynamics of the different shells is given by:

\begin{equation}
\label{eq:sabra}
    \frac{d u_n}{d t} = i \Big(ak_{n+1} u_{n+2} u_{n+1}^* + b k_n u_{n+1} u_{n-1}^* -c k_{n-1} u_{n-1} u_{n-2}\Big) - \nu k_n^2 u_n + f_n\,,
\end{equation}

where $u_n \!\in\! \mathbb{C}$,  represents the velocity fluctuations at a given shell $n$ \red{and $\,^*$ denotes complex conjugation. We set the free parameters $(a,b,c) = (1, -1/2, -1/2)$ such that  $a+b+c=0$, which corresponds to conservation of energy $E \!=\! \sum_n |u_n|^2$ and helicity $H \!=\! \sum_n (a/c)^n|u_n|^2$ in the inviscid, unforced limit.} 

The first term in the RHS of~\autoref{eq:sabra} represents the non-linear term, where we can see that only nearest and next-to-nearest neighbour interactions are preserved, making this model local in Fourier space; the second term represents the viscous term; and the third term represents large-scale forcing. \red{In our simulations, a constant forcing is applied on the first two shells.}

\begin{figure}[htb]
    \centering
    \includegraphics[width=\figSizeSingle\linewidth]{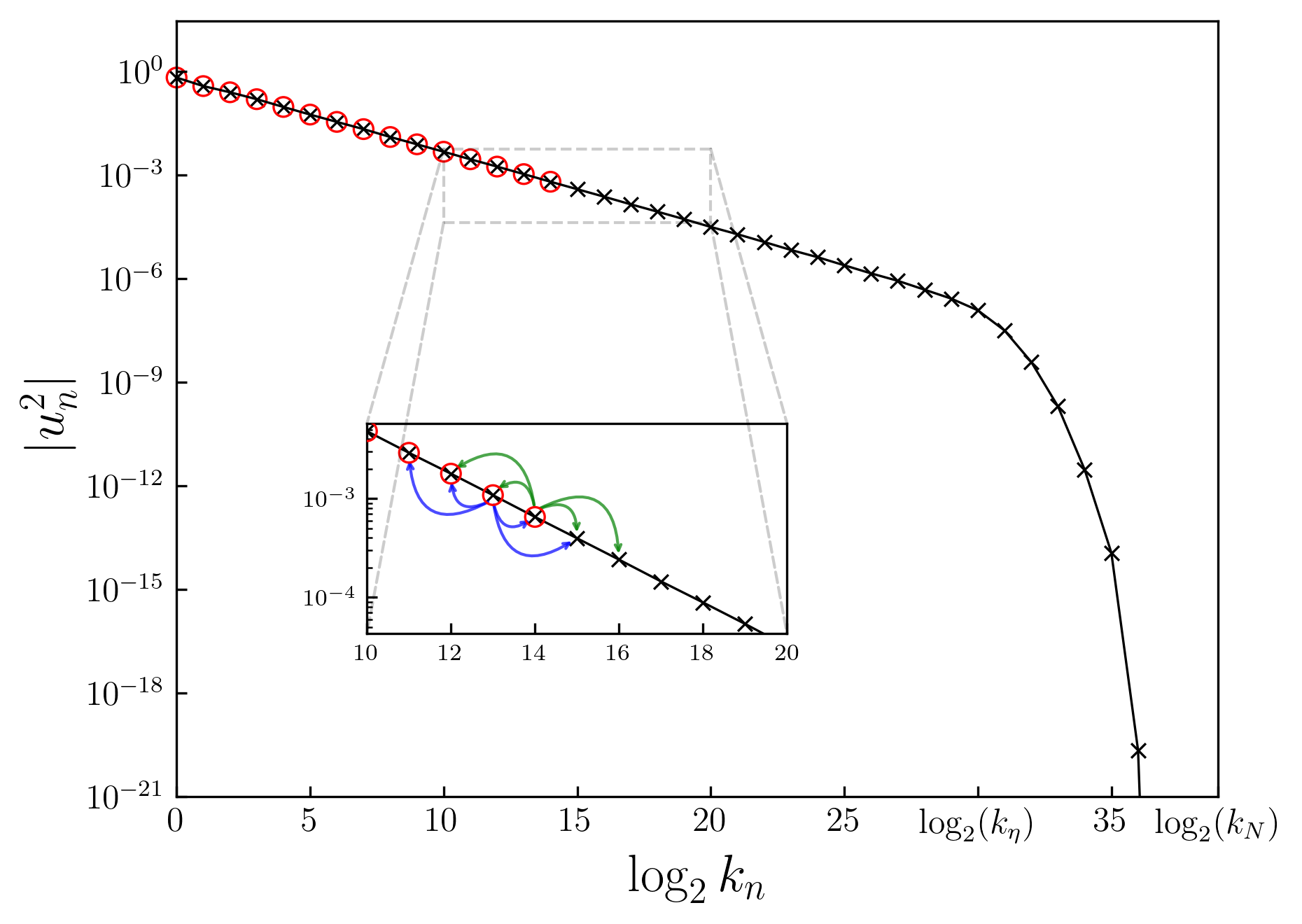} \vspace*{-2mm}
    \caption{Energy spectrum \( |u_n|^2 \) as a function of the wavenumber \( k_n \). The red-circled markers indicate the reduced system, while black crosses represent the fully resolved system. The inset zooms into the cutoff region, highlighting the nonlinear energy transfer through triadic interactions, with blue and green arrows indicating interactions for the two shells preceding the cutoff. It becomes clear that to close the system, only the two shells after the cutoff are needed to model. The x-axis is labeled in \( \log_2 k_n \), along with the Kolmogorov scale \( k_\eta \) and maximum resolved wavenumber \( k_N \).}
    \label{fig:LES}
\end{figure}

The left boundary conditions (BCs) are $u_{-1} \!=\! u_{-2} \!=\! 0$. For the fully resolved system, we integrate~\autoref{eq:sabra} for shells $n \!=\! 0, \ldots, N$, where $N$ is chosen with some margin of the dissipative cutoff scale, $N_{\eta}$, as to fully resolve all of the dynamics. In the case of closing the system in an LES fashion, we have to consider a cutoff shell $N_c$, after which we only have to model the shells $N_c\!+\!1$ and $N_c\!+\!2$ (the left BCs remain the same), as depicted in~\autoref{fig:LES}. \red{In all simulations, we set \(N_c \!=\! 14\), a cutoff chosen within the inertial range but
safely above the dissipative cutoff}. The velocity fields of the subgrid shells then appear explicitly in the equations for $u_{N_c}$ and $u_{N_c-1}$:
\begin{subequations}
\begin{align}
\frac{d u_{N_c}}{d t} &= i \Big(k_{N_c+1} u_{N_c+2} u_{N_c+1}^* - \frac{1}{2} k_{N_c} u_{N_c+1} u_{N_c-1}^* + \frac{1}{2} k_{N_c-1} u_{N_c-1} u_{N_c-2}\Big) - \nu k_{N_c}^2 u_{N_c} \label{eq:2a}, \\
\frac{d u_{N_c-1}}{d t} &= i \Big(k_{N_c} u_{N_c+1} u_{N_c}^* -\frac{1}{2} k_{N_c-1} u_{N_c} u_{N_c-2}^* + \frac{1}{2} k_{N_c-2} u_{N_c-2} u_{N_c-3}\Big) - \nu k_{N_c-1}^2 u_{N_c-1}. \label{eq:2b}
\end{align}
\end{subequations}

%% file: sections_V2/method.tex
\section{\emph{A posteriori} closure}

To close the system, we use a deep neural network that predicts the two missing shells, $u_{N_c+1}$ and  $u_{N_c+2}$, using as input the three preceding shells (enough to fix the flux locally). \red{Unlike an \emph{a priori} approach, which trains the model to match instantaneously the subgrid terms}, we follow an \emph{a posteriori} or \emph{solver-in-the-loop} strategy, where the neural network operates within the evolving simulation. This allows it to learn not only accurate short-term predictions but also how its own errors propagate over time and influence the system’s long-term behavior.

{\color{black}
A key motivation for adopting an \emph{a posteriori} strategy is the presence of distribution shift~\cite{quionero2009dataset}, that is, a mismatch between the statistical distributions encountered during training and deployment. Here, we denote by $x$ the model inputs and by $y$ the corresponding outputs or targets. Three main forms are typically distinguished.
(i)~\emph{Covariate shift}, where the input distribution changes, $p_{\text{train}}(x)\neq p_{\text{test}}(x)$, while the conditional mapping remains the same, $p(y|x)$; 
(ii)~\emph{Prior} or \emph{label shift}, where the output distribution differs, $p_{\text{train}}(y)\neq p_{\text{test}}(y)$, but $p(x|y)$ is preserved; 
and (iii)~\emph{Concept shift}, where the underlying relationship between inputs and outputs itself changes, $p_{\text{train}}(y|x)\neq p_{\text{test}}(y|x)$. 

In our setting, the shift arises as a dynamically induced \emph{covariate shift}. 
During solver deployment, small prediction errors in the learned closure accumulate and perturb the system dynamics, thereby altering the distribution of the model inputs---i.e., the resolved variables provided to the closure at each time step. 
Although the physical relation governing the coupling between resolved and unresolved scales remains invariant, the model gradually encounters regions of phase space that were rare or unseen during training. 
This phenomenon parallels what is observed in autoregressive generative models, such as language models, where compounding errors cause trajectories to drift away from the data manifold~\cite{LLM_dist_shift}. Embedding the neural network directly within the governing equations during training (the solver-in-the-loop strategy) exposes it to its own prediction-induced state distribution, thus reducing the mismatch between training and deployment distributions. 
This mitigates the covariate shift, leading to improved stability, generalization, and a more physically consistent closure.
}

To formalize this approach, let the reduced system $(\tilde{u}_0, \tilde{u}_1, \dots, \tilde{u}_{N_c})$ reside in the \emph{LES manifold} $\mathcal{S}$, while the full system $(u_0, u_1, \ldots, u_N)$ spans the \emph{reference manifold} $\mathcal{R}$. The governing equations of the shell model define a dynamical system on $\mathcal{R}$, but large eddy simulations truncate the system to $\mathcal{S}$, introducing closure errors. To bridge this gap, we introduce a correction operator $\mathcal{C}: \mathcal{S} \!\rightarrow\! \bar{\mathcal{R}}$ parameterized by a neural network, where $\bar{\mathcal{R}}$ represents the truncated reference manifold, restricted to the resolved scales $(u_0, u_1, \ldots, u_{N_c})$. This operator predicts the unresolved shells $ \tilde{u}_{N_c+1}, \tilde{u}_{N_c+2} = \mathcal{C}(\tilde{u}_{N_c-2}, \tilde{u}_{N_c-1}, \tilde{u}_{N_c}) $, enabling reconstruction of the full state in $\mathcal{C}$, via an explicit time integration of the missing terms: this corresponds to an additive correction to the resolved terms that are integrated through a fourth-order Runge-Kutta scheme~\cite{kutta1901beitrag}.   

The operator $\mathcal{C}$ is implemented as a fully connected feedforward neural network (multi-layer perceptron, MLP) with $ L $ hidden layers. Given an input vector  
\[
\bm{x} = (\tilde{u}_{N_c-2}, \tilde{u}_{N_c-1}, \tilde{u}_{N_c}) \in \mathbb{C}^3,
\]
the output $\tilde{\bm{u}} = (\tilde{u}_{N_c+1}, \tilde{u}_{N_c+2}) \in \mathbb{C}^2$ is computed through a sequence of affine transformations and nonlinear activations:
\begin{subequations}
\label{eq:NN_forward}
\begin{align}
\bm{h}^{(0)} &= \bm{x}, \label{eq:NN_forward_a}\\
\bm{h}^{(\ell)} &= \sigma\!\left(\bm{W}^{(\ell)} \bm{h}^{(\ell-1)} + \bm{b}^{(\ell)}\right),
\quad \text{for } \ell = 1, \dots, L, \label{eq:NN_forward_b}\\
\tilde{\bm{u}} &= \bm{W}^{(L+1)} \bm{h}^{(L)} + \bm{b}^{(L+1)}. \label{eq:NN_forward_c}
\end{align}
\end{subequations}

Here, $\bm{W}^{(\ell)}$ and $\bm{b}^{(\ell)}$ are the weight matrix and bias vector for the $\ell$-th layer, while $\sigma(\cdot)$ is a nonlinear activation function, in our case ReLU~\cite{RELU}. The final layer has a linear activation to ensure unrestricted output values.  

\begin{figure}[htb]
    \centering
    \includegraphics[width=1.0\linewidth]{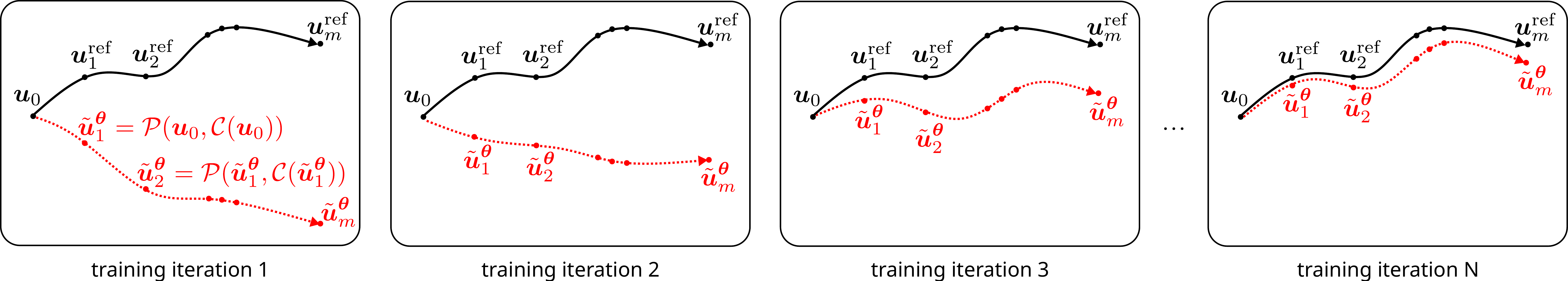}
    \caption{Sketch illustrating the \emph{solver-in-the-loop} approach. The LES trajectory $\tilde{\bm{u}}$ is evolved over $m$ time steps using the discrete solver $\mathcal{P}$, while the correction operator $\mathcal{C}$, parameterized by a neural network, predicts the unresolved scales. The objective is to iteratively refine $\mathcal{C}$ during training, bringing $\tilde{\bm{u}}$ closer to the reference trajectory $\bm{u}^{\text{ref}}$, which integrates the full system dynamics from the same initial condition. Here, the subscript denotes the time step index of the trajectory, i.e., $\bm{u}_k = \bm{u}(t+k\Delta{t})$, and not the shell index.}

    \label{fig:sol-closure}
\end{figure}

The evolution of the system is governed by a discrete solver $\mathcal{P}$, which advances the resolved shells via:  
\begin{equation}
\tilde{\bm{u}}(t+\Delta t) = \mathcal{P}(\tilde{\bm{u}}(t), \tilde{u}_{N_c+1}, \tilde{u}_{N_c+2}),
\end{equation}
where $\tilde{\bm{u}}(t) = (\tilde{u}_0, \dots, \tilde{u}_{N_c})|_t$. Unlike \emph{a priori} closure models trained to minimize \red{instantaneous} errors, our \emph{a posteriori} approach integrates $\mathcal{C}$ recurrently into $\mathcal{P}$. During training, the solver interacts with $\mathcal{C}$ over $m$ steps, exposing it to its own error dynamics. The loss function evaluates deviations from the reference trajectory across this horizon:  
\begin{equation}
\mathcal{L}(\tilde{\bm{u}}^{\bm{\theta}}, \bm{u}^{\text{ref}}) = \sum_{k=1}^m \left\| \tilde{\bm{u}}^{\bm{\theta}}(t+k\Delta t) - \bm{u}^\text{ref}(t+k\Delta t) \right\|_2^2,
\end{equation}
where $\bm{u}^\text{ref}$ is obtained from a direct numerical simulation of $\mathcal{R}$ and $\tilde{\bm{u}}^{\bm{\theta}}$ is the LES solution with the neural network parameters $\bm{\theta} = \{ \bm{W}^{(\ell)}, \bm{b}^{(\ell)} \}_{\ell=1}^{L+1}$. The optimization problem is then formulated as:  
\[
\min_{\bm{\theta}} \mathcal{L}(\tilde{\bm{u}}^{\bm{\theta}}, \bm{u}^{\text{ref}}).
\] 
This multi-step optimization ensures $\mathcal{C}$ learns to mitigate error accumulation and distribution shift, as illustrated in~\autoref{fig:sol-closure}.

{\color{black}
The MLP used for the closure has seven hidden layers of 256 neurons each. Training is performed with the Adam optimiser~\cite{kingma2015adam}, using a learning rate of $10^{-4}$. A detailed summary of the numerical and training parameters as well as the algorithm for training is provided in Appendix~\ref{app:training-details}. For more detailed information regarding the training procedure we refer the readers to Ref.~\cite{freitas25}.
}

%% file: sections_V2/sec3.tex
\section{Results}
In this section, our DL-closed reduced system of equations, hereon referred to as LES-NN, is evaluated by comparing its predictions against the fully resolved ground truth across multiple statistical measures. We analyze its performance in reproducing key turbulence characteristics, including structure functions, temporal evolution of energy, and scale-dependent correlations. Particular attention is given to the ability of the model to capture the tails of joint probability density functions between resolved and unresolved variables; as well as scale invariant statistics following the third hypothesis of K62 theory~\cite{Kolmogorov_1962}, and whether this symmetry is preserved by the closure. \red{All LES-NN results are obtained starting from initial conditions not used during training, and the trajectories are evolved for time horizons substantially longer than those present in the training data (details in Appendix~\ref{app:training-details}). 
Thus, the form of generalization assessed here is in-distribution: robustness to unseen initial conditions and to long-time evolution, rather than out-of-distribution extrapolation across different Reynolds numbers or cutoff scales.}

The structure functions, defined as the difference between velocity increments across different scales to some power $p$, can be defined for the shell models as

\begin{equation}
    S_n^{(p)} = \langle|u_n|^p\rangle,
\end{equation}

\begin{figure}[htb]
    \centering
    \includegraphics[width=\figSizeSingle\linewidth]{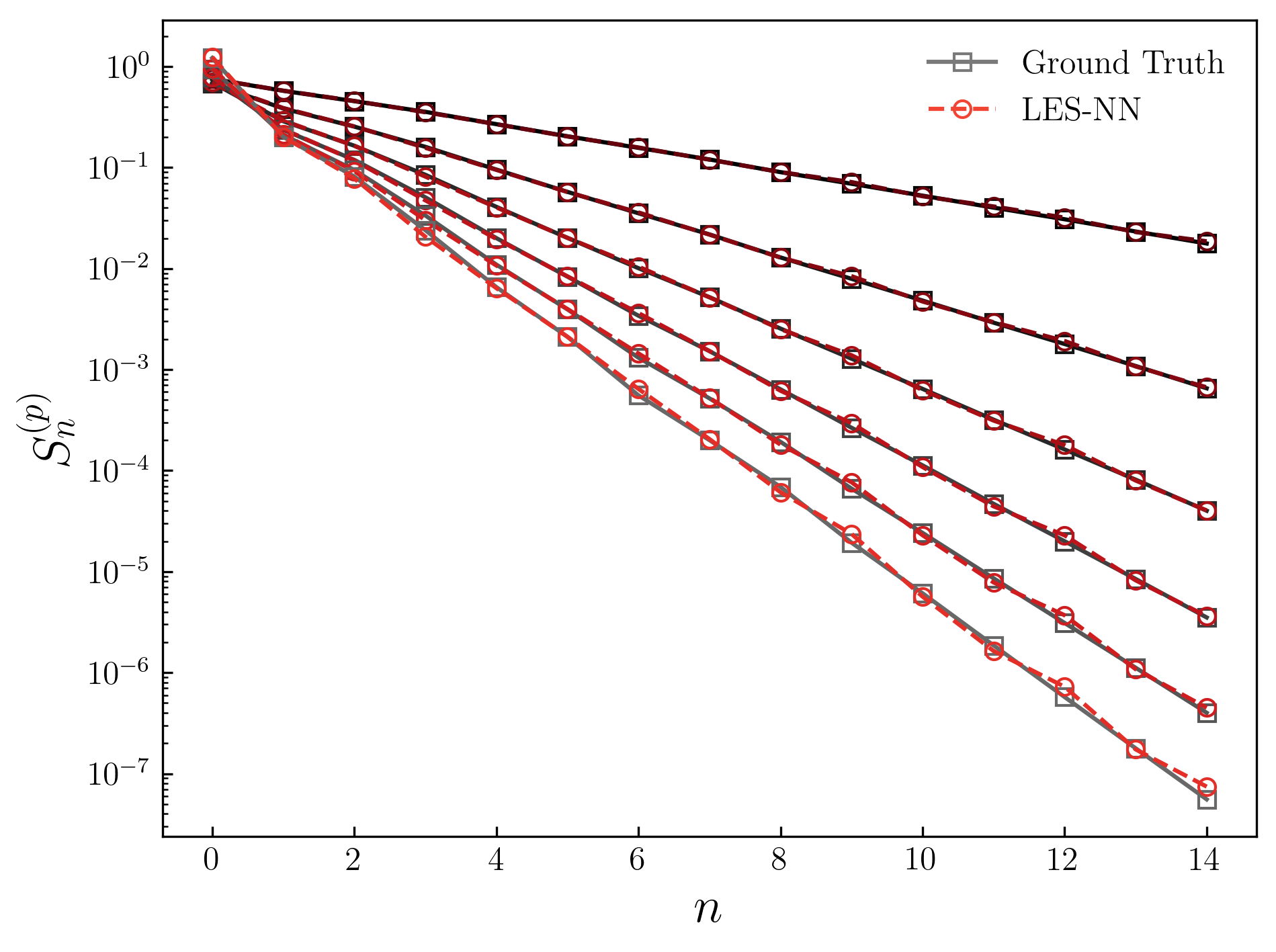}
\caption{Structure functions \( S_n^{(p)} \) as a function of shell index \( n \) for different orders \( p =1, \ldots,6\). The LES-NN model is compared with the ground truth solution.}
    \label{fig:sfs}
\end{figure}

\red{where $\langle\cdot\rangle$ denotes averaging in time and across different initial conditions}. These are shown in~\autoref{fig:sfs} up to the cutoff shell, for $p = 1, \ldots, 6$ for both the fully resolved system (ground truth) and the LES-NN model. There is a great agreement between the model and the ground truth for all orders of structure functions showed, even close to the cutoff scale, \red{which is where the effect of the subgrid scales is more noticeable.}

\begin{figure}[htb]
    \centering
    \includegraphics[width=1.0\linewidth]{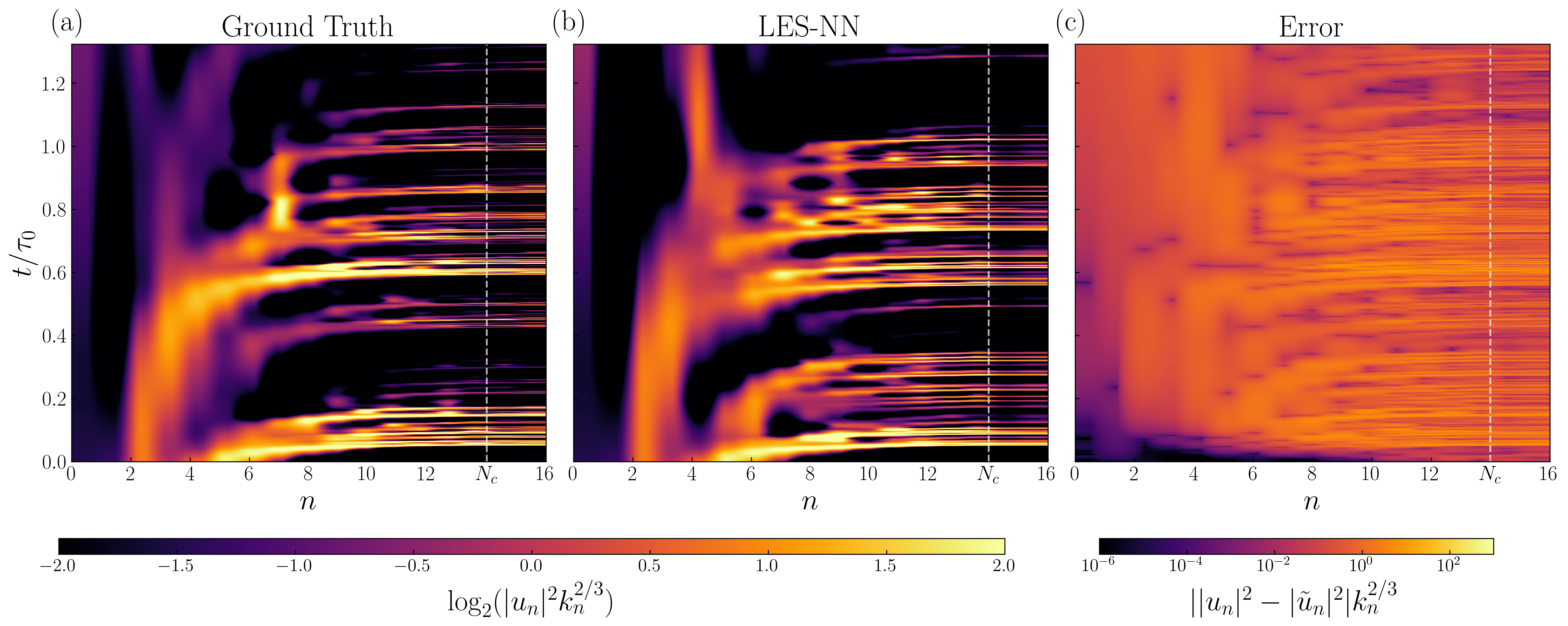}
    \vspace*{-5mm}
    \caption{Time evolution of the normalised energy density \(\log_2(|u_n|^2 k_n^{2/3})\) for (a) the ground truth and (b) the LES-NN model. (c) Absolute normalised error \(\left||u_n|^2 - |\tilde{u}_n|^2\right| k_n^{2/3}\) between the two fields. The x-axis represents the wavenumber index \(n\), \red{while the y-axis represents time normalised by the integral time scale \(t/\tau_0\).} The dashed line represents the cutoff scale.}
    \label{fig:energy_t}
\end{figure}

In~\autoref{fig:energy_t}, we show the time evolution of the logarithm of the \red{normalised energy field} (velocity field scales as $u_n \!\sim\! \smash{k_n^{-1/3}}$) for both the ground truth (GT) and the LES-NN model, along with the normalized error between the two. Both the GT and LES-NN models start from the same initial condition (IC). It is evident that the LES-NN model rapidly decorrelates from the GT. This behavior is expected, as we cannot synchronize past the Lyapunov time. For this particular IC, the discrepancy is particularly clear because we have chosen an IC that is within a burst of energy that has already propagated to the small scales, where the dynamics are much faster. \red{As a result, the effective Lyapunov time for the LES can be roughly approximated as the time scale of the cutoff shell.}

\begin{figure}[htb]
    \centering
    \includegraphics[width=0.9\linewidth]{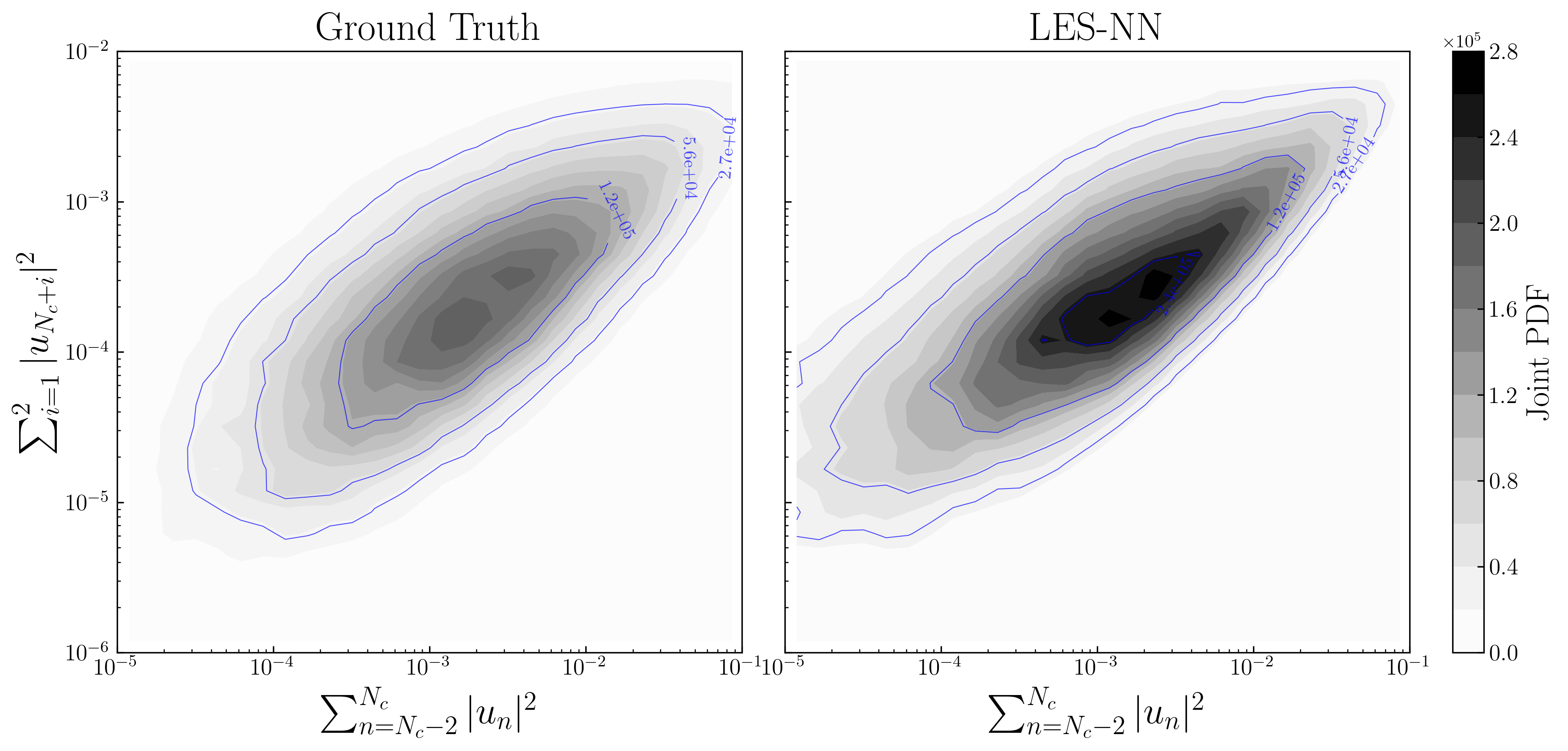} \vspace*{-2mm}
    \caption{Contours of the joint PDF of the summed energy across the input shells for the neural network (x-axis) and the subgrid-scale output shells (y-axis) for both the ground truth and the LES-NN model.}
    \label{fig:energy_corr}
\end{figure}

\red{In order to assess the cross-correlations between resolved and unresolved scales},~\autoref{fig:energy_corr} shows the joint PDF between the summed energy of the \emph{resolved} shells used as input to the neural network, ${\tilde{u}_{N_c}, \tilde{u}_{N_c-1}, \tilde{u}_{N_c-2}}$,  and the sum of the \emph{unresolved} energy, i.e., the sum of the energies of the shells that the NN outputs, ${\tilde{u}^{\theta}_{N_c+1}, \tilde{u}_{N_c+2}^{\theta}}$, for both the ground truth and the LES-NN model. \red{The overall shape of the distribution is well captured, but deviations appear in the joint tails, indicating that the learned closure underestimates the strength of the cross-scale correlations, which are connected with the correlation of the shell phases. We discuss possible causes of this discrepancy at the end of this section.}

In his seminal 1962 study~\cite{Kolmogorov_1962}, Kolmogorov advanced a theoretical framework to characterize the intermittency observed in turbulence at small scales. Central to this framework was the concept of \emph{multipliers}, defined as scale-dependent ratios of velocity increments. These multipliers, denoted as \(w_{ij}(x; \ell, \ell')\), were expressed as:

\begin{equation}
w_{ij}(x; \ell, \ell') = \frac{\delta_i v_j(x, \ell)}{\delta_i v_j(x, \ell')}\,,
\end{equation}

where \(\delta_i v_j(x, \ell) = v_j(x + \ell \mathbf{e}_i) - v_j(x)\) represents the increment of the \(j\)-th velocity component across a spatial separation \(\ell\) along the unit vector \(\mathbf{e}_i\). Kolmogorov posited that at sufficiently high Reynolds numbers, the statistical distributions of these multipliers would exhibit universality--dependent solely on the scale ratio \(\ell/\ell'\) and independent of the absolute scales \(\ell\) or \(\ell'\). Furthermore, he hypothesized statistical independence between multipliers associated with scales separated by large intervals, a conjecture later termed his third hypothesis~\cite{Chen2003_K62_3}. 



For shell models, one can easily define amplitude and phase multipliers~\cite{Benzi93}, respectively, via:
\begin{equation}
    w_n(t) = \left|\frac{u_n(t)}{u_{n-1}(t)}\right|\,,
\end{equation}
\begin{equation}
    \Delta_n(t) = \varphi_{n}(t) - \varphi_{n-1}(t) - \varphi_{n-2}(t)\,,
\end{equation}

where $\varphi_n = \arg u_n$.

\begin{figure}[htb]
    \centering
    \includegraphics[width=0.9\linewidth]{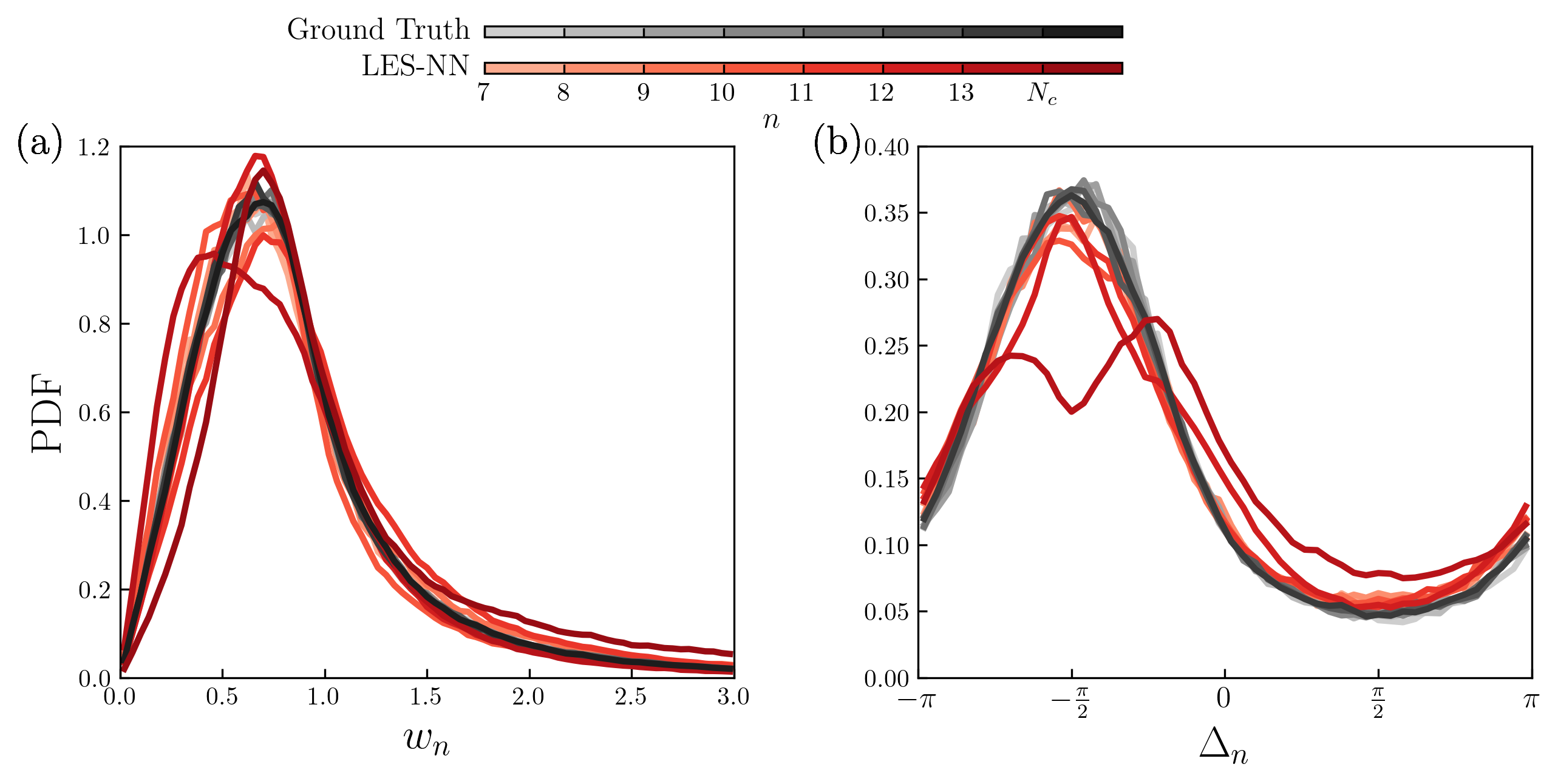}\vspace*{-4mm}
    \caption{PDF of the (a) amplitude multipliers, $w_n$, and (b) phase multipliers, $\Delta_n$, for the ground truth and LES-NN model, for the shells $7$ to $N_c$.}
    \label{fig:multipliers_panel}
\end{figure}

In~\autoref{fig:multipliers_panel}, we show the PDF of the amplitude and phase multipliers for the ground truth and our model, LES-NN, \red{for shells $n \!=\! 7, \ldots, N_c$}. For the fully resolved model, we observe the expected collapse. However, for the LES-NN model, the learned closure breaks this symmetry, and the multipliers are no longer scale invariant near the cutoff. This is a limitation of the model.

\begin{figure}[htb]
    \centering
    \includegraphics[width=0.9\linewidth]{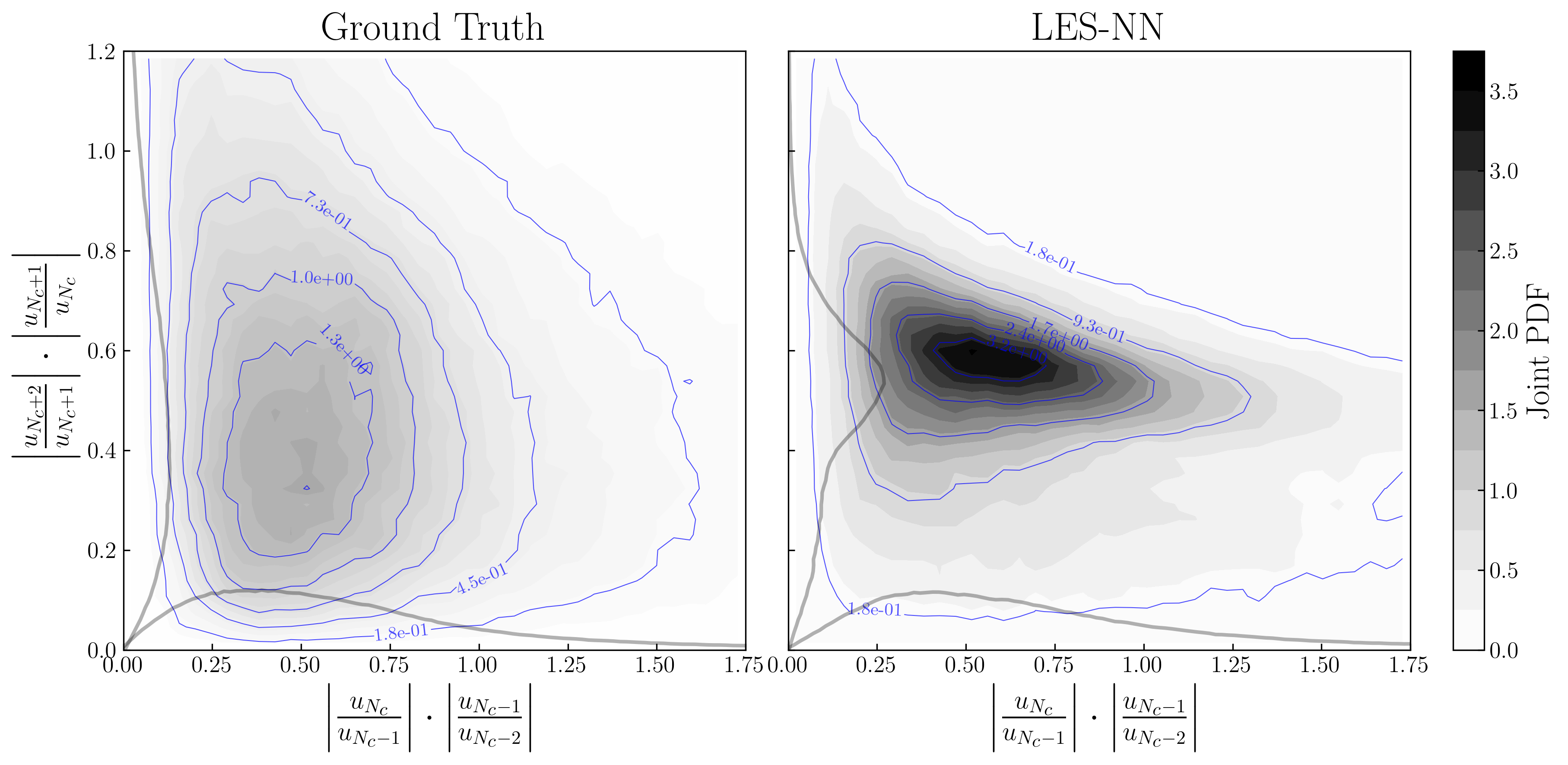}
    \caption{Contours of the joint PDF between the product of the last two resolved multipliers (x-axis) and the product of the two subgrid scale multipliers (y-axis) for both the ground truth and the LES-NN model.}
    \label{fig:joint_product_mult}
\end{figure}

We extend this analysis in~\autoref{fig:joint_product_mult} by examining the joint PDF of the amplitude of the product of the last two resolved multipliers and the product of the two unresolved multipliers (which include the neural network's predictions). The LES-NN model accurately captures the resolved multipliers but fails significantly in the unresolved ones, leading to incorrect predictions not only in the tails but also in the mean. 


{\color{black}
To further probe correlations, following Eyink et al.~\cite{eyink2003gibbsianhypothesisturbulence}, we introduce variables analogous to the spins of the two-dimensional XY model~\cite{Berezinskii:1970pzv, Kosterlitz_1973},
\begin{equation}
    \sigma_n(t)=\ln w_n(t), \qquad U_n(t)=\exp\!\big(i\,\Delta_n(t)\big),
\end{equation}
where $\sigma_n$ quantifies local amplitude fluctuations and $U_n$ encodes relative phase differences between neighboring shells. The set $\{U_n\}$ behaves as interacting rotators on the unit circle, whose alignment measures the degree of phase coherence across scales. In this analogy, the energy cascade corresponds to the propagation of order in a spin chain: strong alignment implies coherent phase dynamics and efficient energy transfer. To quantify this behavior, we compute correlation functions,
\begin{equation}
    C_{XY}(n,m)=\langle X_n^{*}Y_m\rangle-\langle X_n^{*}\rangle\langle Y_m\rangle,
\end{equation}
for $X\!=\!Y\!=\!\sigma$ and $X\!=\!Y\!=\!U$, which measure how amplitude and phase coherence decay with scale separation. The reference shell, $n\!=\!11$, is chosen within the inertial range, close to but below the cutoff. The results are shown in~\autoref{fig:c_xy} for the GT and LES-NN model. We see the expected fast decay with increasing $|m-n|$; however, the correlations near the cutoff are wrongly estimated for the LES-NN model. This helps shine a light on the lack of collapse of the multipliers near the cutoff.

\begin{figure}[htb]
    \centering
    \includegraphics[width=\figSizeSingle\linewidth]{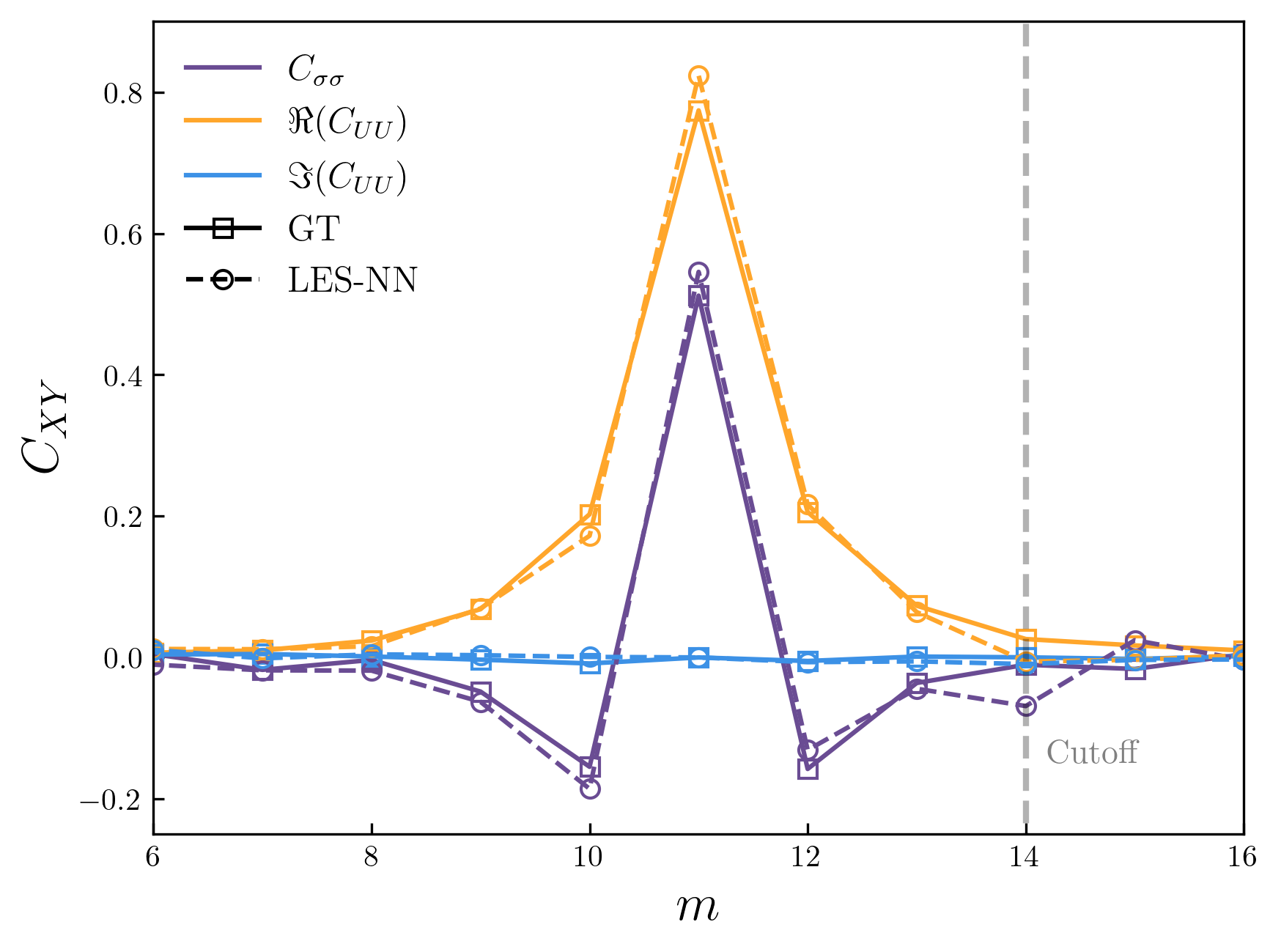}
    \caption{
    {\color{black}
    Correlation functions $C_{XY}(n,m)$ for amplitude fluctuations ($X\!=\!Y\!=\!\sigma$) and phase rotations ($X\!=\!Y\!=\!U$), using shell $n\!=\!11$ as the reference within the inertial range. 
    The curves show how correlations decay as the shell separation $|m-n|$ increases for both the ground truth and the LES--NN model. 
    The dashed vertical line marks the cutoff shell $N_c$. 
    While both models display the expected rapid decay at large separations, the LES--NN correlations deviate near the cutoff, indicating a loss of phase and amplitude coherence in this region.}}
    \label{fig:c_xy}
\end{figure}

The results presented above highlight both the strengths and limitations of the learned closure. The LES-NN model reproduces the resolved-scale statistics with high accuracy, including the overall scaling of the structure functions up to high orders, although small deviations appear at the highest orders. However, these residual differences become more evident when we examine observables that explicitly couple resolved and unresolved variables, such as the joint PDFs and the multipliers near the cutoff. In these quantities, the model underestimates the tails of the distributions and fails to recover the full joint variability between scales. This analysis thus acts as a magnifying glass, revealing the origin of the small but systematic departures observed in the higher-order statistics of the resolved variables.

Several factors could in principle contribute to these discrepancies. One possibility is the neural-network architecture itself: a fully connected MLP may not possess sufficient expressive power to capture all higher-order correlations, and spectral bias could suppress the small-scale variability associated with strong intermittency. 
However, in our tests, changing the network architecture and increasing its depth or width (above what we considered for the results presented here) did not lead to significant improvements. Similarly, changing the loss function to include the subgrid shells explicitly, rather than only the resolved variables, did not alter the results. 
These observations suggest that the origin of the discrepancies does not lie primarily in architectural choices or in the omission of subgrid shells from the loss function.

Instead, our findings point to a structural limitation arising from the formulation of the closure problem itself, particularly the Markovian nature of the model. 
The Mori–Zwanzig (MZ) formalism~\cite{mori,1zwanzig} provides a natural perspective on this issue. When a high-dimensional dynamical system is projected onto a reduced set of resolved variables (as it is done in LES), the resulting exact evolution equation contains two additional contributions beyond the Markovian term: a memory kernel, encoding the delayed interactions between resolved and unresolved variables, and an orthogonal dynamics term that accounts for unresolved-unresolved interactions that cannot be reproduced with the memory term and are usually modeled as noise. A purely Markovian closure omits both effects, and thus misses key ingredients that should in principle help in capturing the correlations between resolved and unresolved variables. This structural lack of memory offers a possible explanation for the breakdown of scale-invariant multiplier statistics near the cutoff and the incomplete reconstruction of resolved-unresolved correlations observed in our results.

Possible ways forward include extending the framework to incorporate non-Markovian effects in the spirit of the MZ formulation, for instance by learning effective memory-kernel or orthogonal-dynamics terms directly from data, as recently explored in data-driven MZ approaches~\cite{regression_mz,dewit2025memoryneeddatadrivenmorizwanzig}. 
In parallel, one may also improve the closure by explicitly incorporating the known scale invariance of multipliers into the learning process, either as a penalty in the loss function or as hard constraint on the model architecture, so that the collapse of the multiplier PDFs is enforced rather than left as an emergent property. 
Another alternative is to reformulate the problem in terms of multipliers themselves by rescaling the governing equations, as done by Domingues Lemos et al.~\cite{JDL24,JDL25} within the hidden-symmetry framework~\cite{HS_21}. 
Combining these ideas---for example, by training a solver-in-the-loop model directly in the multiplier variables or by enforcing multiplier symmetries within the present formulation---may lead to closures that simultaneously reproduce high-order resolved statistics and respect the scale-invariant structure of the multipliers. These considerations are an attempt at clarifying the origin of the observed discrepancies and point to concrete directions for improving the closure framework.
}

%% file: sections_V2/sec4.tex
\section{Conclusion}
{\color{black}

In this work, we have examined in detail a data-driven closure for a shell model of turbulence trained with a solver-in-the-loop (a posteriori) strategy~\cite{freitas25}. 
Building on our previous study, which focused mainly on high-order statistics of the resolved variables, we have gone beyond single-scale observables and analysed quantities that explicitly couple resolved and unresolved degrees of freedom. 
In particular, we investigated joint probability density functions between resolved and subgrid shells, Kolmogorov multipliers and their scale invariance, and correlation functions inspired by the XY-model analogy~\cite{eyink2003gibbsianhypothesisturbulence, Berezinskii:1970pzv, Kosterlitz_1973}. 

Our results show that our data-driven closure reproduces resolved-scale statistics with high accuracy, including structure functions up to order $p\!=\!6$, and remains stable under long-time integration. 
However, when we probe different observables that evaluate the correlations between resolved and unresolved shells, systematic discrepancies emerge. 
The joint PDFs between resolved and unresolved energies reveal an underestimation of the tails, indicating that the model does not fully capture the variability across the cutoff. 
Similarly, the amplitude and phase multipliers lose their expected scale invariance near the cutoff, and the spin–spin correlations involving shells close to the cutoff are slightly misestimated. 
These diagnostics act as a magnifying glass: they expose limitations that are not apparent from the resolved-scale statistics alone.

We have argued that these shortcomings are unlikely to be resolved merely by increasing network capacity or by including subgrid shells directly in the loss function, and instead point to a possible structural limitation of a Markovian closure. 
Within the Mori-Zwanzig formalism~\cite{mori,1zwanzig}, an exact coarse-grained description necessarily involves a memory kernel and an orthogonal dynamics term. 
A memoryless closure, such as the one considered here, cannot fully reproduce these effects, which offers a plausible explanation for the breakdown of multiplier scale invariance near the cutoff and the incomplete reconstruction of resolved-unresolved correlations.

These findings suggest several directions for future work. 
On the one hand, incorporating non-Markovian effects in a principled way, for example via data-driven MZ operators~\cite{regression_mz,dewit2025memoryneeddatadrivenmorizwanzig}, may allow closures that better capture multi-scale correlations across the cutoff. 
On the other hand, explicitly embedding key symmetries of turbulence into the learning problem---such as the scale invariance of multipliers, or working directly in the multiplier/hidden-symmetry formulation~\cite{JDL24,JDL25,HS_21}---could lead to models that are both more universal and less sensitive to the choice of cutoff. 
Extending and testing these ideas in three-dimensional Navier–Stokes turbulence, where similar rescaling procedures can be defined~\cite{AA_rescaling_NSE}, remains an important challenge. 
Overall, our analysis suggests that combining solver-in-the-loop training with a physically informed structure is a promising route towards robust and better-performing data-driven subgrid scale closures.
}

%% file: sections_V2/acknowledgements.tex
\section*{Acknowledgments}
\raggedbottom
The authors benefited from discussions with G. Eyink, A. Mailybaev, M. Sbragaglia. This research was supported by European Union’s HORIZON MSCA Doctoral Networks programme under Grant Agreement No. 101072344, project AQTIVATE (Advanced computing, QuanTum algorIthms and data-driVen Approaches for science, Technology and Engineering), the European Research Council (ERC) under the European Union’s Horizon 2020 research and innovation programme Smart-TURB (Grant Agreement No. 882340), and through an Inria Chair.

%% file: sections_V2/appendix.tex
\section{Training details}\label{app:training-details}

The numerical simulations and solver-in-the-loop training procedure were performed using the parameters listed in Table~\ref{tab:params}. 
The corresponding training loop is summarized in Algorithm~\ref{alg:training}. 
The algorithm represents a single training iteration of the solver-in-the-loop scheme. 
At each time step, the neural network predicts the unresolved shells, $\tilde{u}^{>,\theta}_t$, from the resolved state $\tilde{u}^{<}_t$. 
The resolved dynamics are then advanced in time, and the closure terms $\mathcal{C}_{N_c-1}$ and $\mathcal{C}_{N_c}$, which correspond to the contributions excluded from the direct integration because they depend on the predicted unresolved shells, are subsequently added through explicit integration. This procedure is repeated for $\texttt{msteps}$ iterations, producing trajectories used to evaluate the multi-step loss that measures the deviation from the reference solution. Gradients of the loss with respect to the neural-network parameters are then obtained via backpropagation, and the parameters are updated using gradient-based optimization. All code developed for this work is openly available at \href{https://github.com/andremfreitas/sm_loop}{\texttt{github.com/andremfreitas/sm\_loop}}.

\begin{table}[htb]
    \centering
    \caption{Values of the parameters used in the numerical experiments.}
    \begin{tabular}{llp{10cm}}
        \toprule
        \textbf{Parameter} & \textbf{Value} & \textbf{Description} \\
        \midrule
        $\nu$ & $1 \times 10^{-12}$ & Viscosity \\
        $\mathrm{Re}$ & $\approx 10^{12}$ & Reynolds number \\
        $\epsilon$ & $0.5$ & Forcing amplitude \\
        $N$ & $40$ & Number of shells \\
        $N_\eta$ & $30$ & Kolmogorov scale \\
        $N_\mathrm{c}$ & $14$ & Subgrid cutoff scale \\
        $\tau_0$ & $7.553 \times 10^{-1}$ & Eddy turnover time at the integral scale \\
        $\tau_\eta$ & $1.8367 \times 10^{-6}$ & Eddy turnover time at the dissipative scale \\
        $\Delta t$ & $1 \times 10^{-8}$ & Timestep of ground truth simulation \\
        $\Delta \tilde{t}$ & $1 \times 10^{-5}$ & Timestep of LES–NN model \\
        $N_\mathrm{data}$ & $256$ & Number of initial conditions in the dataset \\
        $N_\mathrm{batch}$ & $1024$ & Batch size used for training \\
        $T_\mathrm{train}$ & $1.65\,\tau_0$ & Integration time of training dataset \\
        $T_\mathrm{test}$ & $3.31\,\tau_0$ & Integration time of test dataset \\
        $\texttt{msteps}$ & $100$ & Time steps in-the-loop per training iteration \\ 
        \bottomrule
    \end{tabular}
    \label{tab:params}
\end{table}

\begin{algorithm}[htb]
\caption{Training loop (single training iteration).}
\begin{algorithmic}[1]
  \State Initialize gradient tape \Comment{Record operations for differentiation}
  \State $\tilde{u} \gets \tilde{u}_0$ \Comment{Batch of initial conditions sampled from dataset}
  \For{$t = 0$ to $msteps-1$}
      \State $\tilde{u}^{>,\theta}_t \gets \text{NN}_{\theta}(\tilde{u}^{<}_t)$ \Comment{Predict unresolved shells via neural network}
      \State $\tilde{u}^{<}_{t+1} \gets \text{RK4}(\tilde{u}^{<}_t)$ \Comment{Advance resolved shells with fourth-order Runge–Kutta}
      \State $\mathcal{C}_{N_c-1} \gets \Delta \tilde{t}\, i(a k_{N_c} \tilde{u}^{\theta}_{N_c+1} \tilde{u}^*_{N_c})$ \Comment{Compute closure term for shell $N_c-1$}
      \State $\mathcal{C}_{N_c} \gets \Delta \tilde{t}\, i(a k_{N_c+1} \tilde{u}^{\theta}_{N_c+2} \tilde{u}^{*,\theta}_{N_c+1} + b k_{N_c} \tilde{u}^{\theta}_{N_c+1} \tilde{u}^*_{N_c-1})$ \Comment{Compute closure term for shell $N_c$}
      \State $\mathcal{C} \gets \text{concatenate}(\mathcal{C}_{N_c-1}, \mathcal{C}_{N_c})$ \Comment{Aggregate closure corrections}
      \State $\tilde{u}^{<}_{t+1} \gets \tilde{u}^{<}_{t+1} + \mathcal{C}$ \Comment{Update resolved shells with closure contribution}
  \EndFor
  \State Compute loss $\mathcal{L}$ \Comment{Compare predicted and reference trajectories}
  \State Compute gradients $\nabla_\theta \mathcal{L}$ \Comment{Differentiate loss w.r.t.\ network parameters}
  \State Update parameters $\theta$ \Comment{Apply optimizer step}
\end{algorithmic}
\label{alg:training}
\end{algorithm}